\documentclass[conference]{IEEEtran}
\usepackage[boxruled]{algorithm2e}
\usepackage{cite}
\usepackage{graphicx}
\usepackage{psfrag}
\usepackage{subfigure}
\usepackage{url}
\usepackage{amsmath}
\usepackage{array}
\usepackage{amssymb}
\usepackage{amsfonts}
\usepackage{graphicx}
\usepackage{epstopdf}
\usepackage{algorithmic}

\newtheorem{lemma}{Lemma}

\newtheorem{definition}{Definition}
\newtheorem{theorem}{Theorem}

\newtheorem{example}{Example}
\newtheorem{note}{Note}
\usepackage{setspace}
\usepackage{amscd}
\usepackage{mathrsfs}
\usepackage{epsfig}
\usepackage{color}
\usepackage{textcomp}
\usepackage{multirow}
\usepackage{threeparttable}
\input{epsf.sty}
\title{Performance Analysis of Adaptive Physical Layer Network Coding for Wireless Two-way Relaying}
\begin{document}

\author{
\authorblockN{Vijayvaradharaj T. Muralidharan and B. Sundar Rajan}
\authorblockA{Dept. of ECE, IISc, Bangalore 560012, India, Email:{$\lbrace$tmvijay, bsrajan$\rbrace$}@ece.iisc.ernet.in
}
}

\maketitle
% \thispagestyle{empty}	
%%%%%%%%
\begin{abstract}
The analysis of modulation schemes for the physical layer network-coded two way relaying scenario is presented which employs two phases: Multiple access (MA) phase and Broadcast (BC) phase. It was shown by Koike-Akino et. al. that adaptively changing the network coding map used at the relay according to the channel conditions greatly reduces the impact of multiple access interference which occurs at the relay during the MA phase. Depending on the signal set used at the end nodes, the minimum distance of the effective constellation at the relay becomes zero for a finite number of channel fade states referred as the singular fade states. %In \cite{NVR}, a method to obtain adaptive network coding maps which remove the harmful effect of most of these singular fade states, which depend on the signal set used, referred as the removable singular fade states, using the completion of partially filled Latin Squares was proposed. The other singular fade states which occur due to channel outage are referred as the non-removable singular fade spaces. With every singular fade state, we can associate an error event during the MA phase. 
The singular fade states fall into the following two classes: The ones which are caused due to channel outage and whose harmful effect cannot be mitigated by adaptive network coding are referred as the \textit{non-removable singular fade states}. The ones which occur due to the choice of the signal set and whose harmful effects can be removed by a proper choice of the adaptive network coding map are referred as the \textit{removable} singular fade states. In this paper, we derive an upper bound on the average end-to-end Symbol Error Rate (SER), with and without adaptive network coding at the relay, for a Rician fading scenario. It is shown that without adaptive network coding, at high Signal to Noise Ratio (SNR), the contribution to the end-to-end SER comes from the following error events which fall as $\text{SNR}^{-1}$: the error events associated with the removable singular fade states, the error events associated with the non-removable singular fade states and the error event during the BC phase. In contrast, for the adaptive network coding scheme, the error events associated with the removable singular fade states contributing to the average end-to-end SER fall as $\text{SNR}^{-2}$ and as a result the adaptive network coding scheme provides a coding gain over the case when adaptive network coding is not used. It is shown that for a Rician fading channel, the error during the MA phase dominates over the error during the BC phase. Hence, adaptive network coding, which improves the performance during the MA phase provides more gain in a Rician fading scenario than in a Rayleigh fading scenario. Also, it is shown that for large Rician factors, among those removable singular fade states which have the same magnitude, those which have the least absolute value of the phase angle alone contribute dominantly to the end-to-end SER and it is sufficient to remove the effect of only such singular fade states.%This reduces the total number of network coding maps used at the relay and hence results in a reduction in the system complexity at the relay as well as the number of overhead bits required for indicating the choice of the adaptive network coding map.  

\end{abstract}
%%%%%%%%%%%%%%%%%%%%%%%%%%%%%%%%
\section{Background and Preliminaries}
\begin{figure}[htbp]
\centering
\subfigure[MA Phase]{
\includegraphics[totalheight=.75in,width=1.5in]{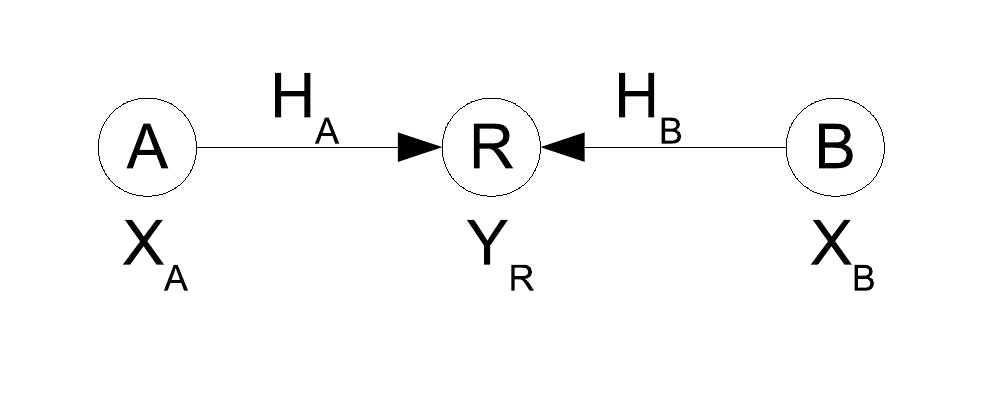}
\label{fig:phase1}
}
\subfigure[BC Phase]{
\includegraphics[totalheight=.75in,width=1.5in]{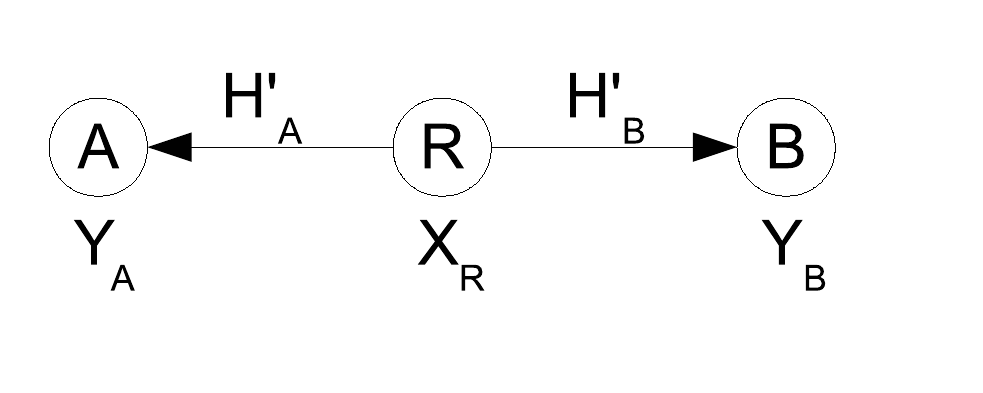}
\label{fig:phase2}
}
\caption{The Two Way Relay Channel}
\label{relay_channel}
\end{figure}

  The wireless two-way relay channel (Fig. \ref{relay_channel}) in which  bidirectional data transfer takes place between the nodes A and B with the help of the relay R is considered. All the three nodes are assumed to have half-duplex constraint, i.e., they cannot transmit and receive simultaneously in the same frequency band. The Denoise and Forward protocol, introduced in \cite{PoYo_DNF}, is considered which consists of the following two phases: the \textit{multiple access} (MA) phase, during which A and B simultaneously transmit to R and the \textit{broadcast} (BC) phase during which R transmits to A and B. Network coding map, also called the denoising map, is employed at R in such a way that A (B) can decode the message of B (A), given that A (B) knows its own message. 
%%%%%%%%%%%%%%%%%%%%%%%%%%%%%%%%%%%
\vspace{-0.1 cm}
%%%%%%%%%%%%%%%%%%%%%%%%%%%%%%%%% 
%%%%%%%%%%%%%%%%%%%%%%%%%%%%%%% 
\subsection{Background}

 The concept of physical layer network coding has attracted a lot of attention in recent times. The idea of physical layer network coding for the two way relay channel was first introduced in \cite{ZLL}, where the multiple access interference occurring at the relay was exploited so that the communication between the end nodes can be done using a two phase protocol. Information theoretic studies for the physical layer network coding scenario were reported in \cite{KMT},\cite{PoY}. A differential modulation scheme with analog network coding for bi-directional relaying was proposed in \cite{LiYoAnBiAt}. The design principles governing the choice of modulation schemes to be used at the nodes for uncoded transmission were studied in \cite{APT1}. An extension for the case when the nodes use convolutional codes was done in \cite{APT2}. A multi-level coding scheme for the two-way relaying scenario was proposed in \cite{HeN}. Power allocation strategies and lattice based coding schemes for bi-directional relaying were proposed in \cite{WiNa}.

 Error analysis of the two-way AWGN relay channel with physical layer network coding based on the Detect and Forward (DF) protocol, in which the relay R transmits the estimate of the Exclusive OR (XOR) of A's and B's transmission bits, was done in \cite{LuFuQiCh}. 
Performance analysis for the two-way Rayleigh fading relay channel with physical layer network coding, based on the Amplify and Forward protocol was presented in \cite{LoLiVu}. For a two-way Rayleigh fading relay channel with BPSK modulation, upper and lower bounds on the Symbol Error Rate (SER) for the DF protocol were obtained in \cite{JuKi}. Exact BER analysis of the two-way Rayleigh fading relay channel with BPSK modulation for the DF protocol was done in \cite{PaChLe}.

While for BPSK modulation, the XOR network code offers the best performance, for other signal sets, changing the network coding map adaptively according to channel conditions provides a significant performance improvement \cite{APT1}. A computer search algorithm called the \textit{Closest-Neighbour Clustering} (CNC) algorithm was proposed in \cite{APT1} to obtain such adaptive network coding maps resulting in the best distance profile at R.  An alternative procedure to obtain the adaptive network coding maps, based on the removal of deep channel fade conditions called singular fade states using Latin Squares was proposed in \cite{NVR}. A quantization of the set of all possible channel realizations based on the network code used was obtained analytically in \cite{VNR}. For the adaptive network coding schemes, performance improvement results due to a proper choice of adaptive network coding maps which effectively mitigate the effect of most of the singular fade states, referred as the removable singular fade states. In \cite{NVR}, a method to obtain adaptive network coding maps which remove the harmful effect of these removable singular fade states, using the completion of partially filled Latin Squares was proposed. 

Unlike the DF protocol, in which XOR map is used irrespective of channel conditions, the average SER analysis of the adaptive network coding schemes proposed in \cite{APT1}, \cite{NVR}-\cite{VNR} should take into account the fact that the network coding maps used depend on the channel fade coefficients. 

In this paper, we derive an upper bound on the end-to-end SER for the adaptive network coding schemes (\cite{APT1},\cite{NVR}-\cite{VNR}) as well as for the case when adaptive network coding is not used, for a Rician fading scenario. With every singular fade state, we can associate an error event during the MA phase. It is shown that without adaptive network coding, at high Signal to Noise Ratio (SNR), the contribution to the end-to-end SER comes from the following error events which fall as $\text{SNR}^{-1}$: the error events associated with the removable singular fade states, the error events associated with the non-removable singular fade states and the error event during the BC phase. In contrast, for the adaptive network coding schemes proposed in \cite{APT1} and \cite{NVR}-\cite{VNR}, the error events associated with the removable singular fade states fall as $\text{SNR}^{-2}$ and as a result the adaptive network coding schemes provide a coding gain over the case when adaptive network coding is not used. Also, it is shown that for sufficiently large Rician factors, only some of the removable singular fade states contribute dominantly to the end-to-end SER and it is sufficient to remove only such singular fade states. %This reduces the total number of network coding maps used at the relay and hence results in a reduction in the system complexity at the relay as well as the number of overhead bits required for indicating the choice of the adaptive network coding map.    
%
%%%%%%%%%%%%%%%%%%%%%%%%%%%%%%%%%%%%%%%%%%%%%%%%%%
\subsection{Signal Model}

Let $\mathcal{S}$ denote the signal set of unit energy used at A and B, with $M=2^\lambda$ points, $\lambda$ being a positive integer. Assume that A (B) wants to transmit a $\lambda$-bit binary tuple to B (A). Let $\mu: \mathbb{F}_2^\lambda \rightarrow \mathcal{S}$ denote the mapping from bits to complex symbols used at A and B. Throughout the paper all the fading coefficients are assumed to be Rician distributed with a Rician factor $K.$ A Rician distributed random variable $X$ has a scattered component and a line of sight component, i.e., $X$ can be written as $\frac{1}{\sqrt{K+1}}X_G+\sqrt{\frac{K}{K+1}}e^{j \theta},$ where $X_G$ is a circularly symmetric complex Gaussian randonm variable with unit variance. Since $\theta$ is a constant, it can be cancelled out at the transmitting nodes. Hence, without loss of generality, we assume $\theta=0.$

\subsubsection*{Multiple Access (MA) phase}
It is assumed that the CSI is not available at the transmitting nodes A and B during the MA phase. Also, a block fading scenario is assumed. Let $x_A= \mu(s_A)$, $x_B=\mu(s_B)$ $\in \mathcal{S}$ denote the complex symbols transmitted by A and B respectively, where $s_A,s_B \in \mathbb{F}_2^\lambda$. The received signal at $R$ is given by,
\begin{align}
\nonumber
Y_R=H_{A} \sqrt{E_s} x_A + H_{B} \sqrt{E_s} x_B +Z_R,
\end{align}
where $H_A$ and $H_B$ are the fading coefficients associated with the A-R and B-R links respectively, which follow Rician distribution with a Rician factor $K.$ Note that the Rician factor is the power ratio between the line of sight and scattered components. The additive noise $Z_R$ is assumed to be $\mathcal{CN}(0,\sigma^2)$, where $\mathcal{CN}(0,\sigma^2)$ denotes the circularly symmetric complex Gaussian random variable with variance $\sigma ^2$. The average energy of A's and B's transmission is equal to $E_s.$ Throughout, by SNR we mean the ratio $\frac{E_s}{\sigma ^2}.$ The ratio $ H_{B}/H_{A}$ denoted as $z=\gamma e^{j \theta}$, where $\gamma \in \mathbb{R}^+$ and $-\pi \leq \theta < \pi,$ is referred as the {\it fade state} and for simplicity, also denoted by $(\gamma, \theta).$ 

Let $\mathcal{S}_{R}(H_A,H_B)$ denote the effective constellation seen at the relay during the MA phase, i.e., 
\begin{align} 
\nonumber
 \mathcal{S}_{R}(H_A,H_B)=\left\lbrace H_A x_A+ H_B x_B \vert x_A,x_B \in \mathcal{S}\right \rbrace.
 \end{align}

Let $d_{min}(H_A,H_B)$ denote the minimum distance between the points in the constellation $\mathcal{S}_{R}(H_A,H_B)$, i.e.,

{\footnotesize
\begin{align}
\label{eqn_dmin} 
d_{min}(H_A,H_B)=\hspace{-0.5 cm}\min_{\substack {{(x_A,x_B),(x'_A,x'_B) \in \mathcal{S}^2} \\ {(x_A,x_B) \neq (x'_A,x'_B)}}}\hspace{-0.5 cm}\vert H_A\left(x_A-x'_A\right)+H_B \left(x_B-x'_B\right)\vert.
\end{align}
}
 
 Let $\Delta \mathcal{S}$ denote the difference constellation of the signal set $\mathcal{S},$ i.e., $\Delta\mathcal{S}=\lbrace x-x' \vert x, x'\in \mathcal{S}\rbrace.$  

 From \eqref{eqn_dmin}, it is clear that there exists values of $(H_A,H_B)$ for which  $d_{min}(H_A,H_B)=0.$
Whether for a given realization of $(H_A,H_B)$, $d_{min}(H_A,H_B)$ is zero or not depends only on the fade state $\gamma e^{j\theta}=\frac{H_B}{H_A}.$ The values of $\gamma e^{j \theta}$ for which $d_{min}(H_A,H_B)=0$ are of the form $-\frac{\Delta x_A}{\Delta x_B},$ where $\Delta x_A=x_A-x'_A,\Delta x_B= x_B-x'_B \in \Delta \mathcal{S}$ and are referred to as the singular fade states \cite{NVR}. Note that $0$ and $\infty$ are also singular fade states which occur when $H_B=0$ or $H_A=0.$ Let $\mathcal{H}$ denote the set of all singular fade states excluding 0 and $\infty.$%Throughout the paper, unless explicitly mentioned, by singular fade state we refer to the ones excluding $0$ and $\infty.$ %Let $\mathcal{H}$ denote the set of all such singular fade states.

Let $(\hat{x}^R_A,\hat{x}^R_B) \in \mathcal{S}^2$ denote the Maximum Likelihood (ML) estimate of $({x}_A,{x}_B)$ at R based on the received complex number $Y_{R}$, i.e.,
 \begin{align}
 (\hat{x}^R_A,\hat{x}^R_B)=\arg\min_{({x}'_A,{x}'_B) \in \mathcal{S}^2} \vert Y_R-H_{A}{x}'_A-H_{B}{x}'_B\vert.
 \end{align}
 
%%%%%%%%%%%%%%%%%%%%%%%%%%%%%%%%%%%%%%%%%%%%%%
\subsubsection*{Broadcast (BC) phase}

Depending on the value of $\gamma e^{j \theta}$, R chooses a many-to-one map $\mathcal{M}^{\gamma,\theta}:\mathcal{S}^2 \rightarrow \mathcal{S}'$, where $\mathcal{S}'$ is the signal set (of size between $M$ and $M^2$) used by R during the $BC$ phase. The elements in $\mathcal{S}^2$ which are mapped on to the same complex number in $\mathcal{S}'$ by the map $\mathcal{M}^{\gamma,\theta}$ are said to form a cluster. Let $\lbrace \mathcal{L}_1, \mathcal{L}_2,...,\mathcal{L}_l\rbrace$ denote the set of all such clusters. The formation of clusters for $\gamma^{j \theta}$ is called clustering, and is denoted by $\mathcal{C}^{\gamma,\theta}$. For examples of clusterings for 4-PSK and 8-PSK signal sets, see \cite{NVR}.

For a given realization of $H_A$ and $H_B,$ the choice of the network coding map depends only on the fade state $\gamma e^{j\theta}=\frac{H_B}{H_A},$ since all the distances between the points in the constellation $\mathcal{S}_{R}(H_A,H_B)$ can be normalized by $H_A$ to make the set of all such distances a function of only $\gamma e^{j \theta}.$ 
 
The received signals at A and B during the BC phase are respectively given by,
\begin{align}
Y_A=H'_{A} \sqrt{E_s} X_R + Z_A,\;Y_B=\sqrt{E_s} H'_{B} X_R + Z_B,
\end{align}
where $X_R=\mathcal{M}^{\gamma,\theta}(\hat{x}_A,\hat{x}_B) \in \mathcal{S'}$ is the complex number transmitted by R. The fading coefficients corresponding to the R-A and R-B links, denoted by $H'_{A}$ and $H'_{B}$ respectively are Rician distributed with Rician factor $K.$ The additive noises $Z_A$ and $Z_B$ are $\mathcal{CN}(0,\sigma ^2$).
In order to ensure that A (B) is able to decode B's (A's) message, the clustering $\mathcal{C}^{\gamma,\theta}$ should satisfy the exclusive law \cite{APT1}, i.e.,

{\small
\begin{align}
\left.
\begin{array}{ll}
\nonumber
\mathcal{M}^{\gamma,\theta}(x_A,x_B) \neq \mathcal{M}^{\gamma,\theta}(x'_A,x_B), \; \mathrm{for} \;x_A \neq x'_A, \; \forall \;x_B \in  \mathcal{S},\\
\nonumber
\mathcal{M}^{\gamma,\theta}(x_A,x_B) \neq \mathcal{M}^{\gamma,\theta}(x_A,x'_B), \; \mathrm{for} \;x_B \neq x'_B, \; \forall \;x_A \in \mathcal{S}.
\end {array}
\right\} 
\end{align}
\vspace{-.3 cm}
}

The node A (B) can decode the message of B (A) by observing $Y_A$ ($Y_B$) through ML decoding, since A (B) knows $x_A$ ($x_B$) and the map $\mathcal{M}^{\gamma,\theta}$ satisfies the exclusive law.

\begin{definition} {\cite{NVR}}
The cluster distance between a pair of clusters $\mathcal{L}_i$ and $\mathcal{L}_j, i\neq j,$ is the minimum among all the distances calculated between the points $H_A x_A+H_B x_B$ and $H_A x'_A+H_B x'_B \in \mathcal{S}_R(H_A,H_B)$, where $(x_A,x_B) \in \mathcal{L}_i$ and $(x'_A,x'_B) \in \mathcal{L}_j$. The \textit{minimum cluster distance} of the clustering $\mathcal{C}^{\gamma,\theta}$ is the minimum among all the cluster distances.
\end{definition}

A clustering $\mathcal{C}^{\lbrace h \rbrace}$ is said to remove a singular fade state $ h \in \mathcal{H}$, if the minimum cluster distance $d_{min}(\mathcal{C}^{\lbrace h\rbrace})$ is greater than zero.

The CNC algorithm proposed in \cite{APT1} obtains the map $\mathcal{M}^{\gamma,\theta}$ which results in the best distance profile during the MA phase at R, for a given $\gamma e^{j\theta}.$ The CNC algorithm is run for all possible channel realizations and a partition of the set of all channel realizations is obtained depending on the chosen network coding map. For a given channel realization, the choice of the network coding map is indicated to A and B using overhead bits. 

The CNC algorithm optimizes the entire distance profile instead of maximizing only the minimum distance. In some cases, this results in the use of signal sets with a larger cardinality during the BC phase and also results in an extremely large number of maps. For instance, for 16 QAM, the CNC algorithm results in more than 18,000 maps \cite{APT1}. To solve this problem, an algorithm called the Nearest Neighbour Clustering (NNC) algorithm was proposed in \cite{APT1} which maximizes the minimum distance alone, instead of optimizing the entire distance profile. 

 In \cite{NVR}, the equivalence between the network coding maps satisfying the exclusive law and the mathematical structure called Latin Squares was shown. Network coding maps which remove all the singular fade states were obtained by the partial completion of Latin Squares.

Consider a singular fade state $h=-\frac{\Delta x_A}{\Delta x_B} \in \mathcal{H},$ where $\Delta x_A$ and $\Delta x_B$ are non-zero complex numbers which belong to $\Delta \mathcal{S}.$ Let $\Delta x_A=x_A-x'_A$ and $\Delta x_B=x_B-x'_B.$ Associated with the singular fade state $h,$ we have the error event that the pair $(x_A,x_B)$ is wrongly decoded as $(x'_A,x'_B).$ The CNC algorithm as well as the scheme proposed in \cite{NVR} remove the singular fade state  $h$ by placing all such pairs $(x_A,x_B)$ and $(x'_B,x'_B)$ for which $\Delta x_A=x_A-x'_A$ and $\Delta x_B=x_B-x'_B$ in the same cluster. 

The harmful effect of the singular fade states 0 and $\infty$ cannot be removed since the pairs $(x_A,x_B)$ and $(x_A,x'_B)$ (and also $(x_A,x_B)$ and $(x'_A,x_B)$) result in these singular fade states and they cannot be placed in the same cluster without violating the exclusive law \cite{APT1}. The singular fade states 0 and $\infty$ which occur due to channel outage, irrespective of the signal set used, are referred as the \textit{non-removable singular fade states}. The rest of the singular fade states which depend on the signal set used are referred as the \textit{removable singular fade states}. Henceforth, unless explicitly mentioned, by singular fade state, we refer to only the removable ones.%Let $\mathcal{H}$ denote the set of all such singular fade states.

Throughout the paper, in a statement if it is mentioned simply as adaptive network coding,  it refers to the scheme proposed in \cite{APT1} as well the one proposed in \cite{NVR}, i.e., the claim made in the statement holds for both the schemes.

In this paper, an upper bound on the average end-to-end SER is obtained for the two-way relaying scenario with and without adaptive network coding. From the obtained analysis, the reason why adaptive network coding provides performance improvement becomes very clear. Also, it is shown that not all the singular fade states contribute equally to the end-to-end error probability. Removing only those singular fade states which contribute dominantly to the end-to-end error probability reduces the system complexity at the relay as well as the number of overhead bits required for indicating the choice of the adaptive network coding map.

The contributions and organization of this paper are as follows:
\begin{itemize}
\item
An upper bound on the average end-to-end SER for the wireless two-way relaying scenario with and without adaptive network coding is derived (Section II A and Section II B).  It is shown that without adaptive network coding, at high SNR, the contribution to the average end-to-end SER comes from the following terms which decrease as $\text{SNR}^{-1}$: the error events associated with the removable and the non-removable singular fade states and the error event during the BC phase. In contrast, for the adaptive network coding schemes proposed in \cite{APT1} and \cite{NVR}, the error events associated with the removable singular fade states fall as $\text{SNR}^{-2}$ and as a result the adaptive network coding schemes provide a coding gain over the case when adaptive network coding is not used.% Also it is shown that as the Rician factor increases, error during the MA phase contributes predominantly to the end-to-end SER. In other words,for high Rician factors, maximizing the minimum cluster distance is more important than maximizing the minimum distance of the signal set used during the BC phase (Section II).
\item
It is shown that in a Rician fading scenario, the error during the MA phase dominates over the error during the BC phase. Hence, the adaptive network coding schemes, which improve the performance during the MA phase, provides more gain in a Rician fading scenario than in a Rayleigh fading scenario (Section II).
\item
It is shown that in a Rician fading scenario, the removal of the singular fade state $s=1$ assumes greatest significance. While in a Rayleigh fading scenario, all the singular fade states contribute dominantly to the overall average SER, it is shown that in a Rician fading scenario, for sufficiently large Rician factors, among those singular fade states which have the same magnitude, only those for which the absolute value of the phase angle is the least contribute dominantly to the end-to-end SER and it is sufficient to remove only such singular fade states (Section III). 
\item  
Simulation results which confirm the above mentioned facts are presented in Section IV.
\end{itemize}

\textbf{\textit{Notations}:}
Q[.] denotes the tail probability of the standard Normal distribution. $P_H{\lbrace E \rbrace}$ denotes the probability of the event $E$ conditioning on the set of random variables $H.$  $\mathbb{E}(X)$ denotes the expectation of $X.$ 
\section{ERROR ANALYSIS OF THE WIRELESS TWO-WAY RELAYING SCENARIO}
In the section, upper bounds on the end-to-end SER are obtained for the wireless two-way relaying scenario with and without adaptive network coding.

Let $\hat{x}_B^{A}$ and $\hat{x}_A^{B}$denote the messages decoded by A and B respectively at the end of the BC phase. Let $E_A$ and $E_B$ respectively denote the error events $\hat{x}_B^{A} \neq x_B$ and $\hat{x}_A^{B} \neq x_A.$ Let $H=(H_A,H_B,H'_A,H'_B)$ denote a particular realization of the channel fade coefficients. The end-to-end SER ${P}_H\lbrace E_A \cup E_B\rbrace$ given in \eqref{pe_avg}, can be upper-bounded as in \eqref{pe_ub} (both shown at the top of the next page).

\begin{figure*}
{\scriptsize
\begin{align}
\nonumber
{P}_H\lbrace E_A \cup E_B\rbrace &= {P}_H \lbrace E_A \cup E_B | \mathcal{M}^{\gamma,\theta}(\hat{x}^R_A,\hat{x}^R_B) = \mathcal{M}^{\gamma,\theta}(x_A,x_B)\rbrace {P}_H \lbrace \mathcal{M}^{\gamma,\theta}(\hat{x}^R_A,\hat{x}^R_B) = \mathcal{M}^{\gamma,\theta}(x_A,x_B)\rbrace\\
\label{pe_avg}
&\hspace{4.6 cm}
+{P}_H \lbrace E_A \cup E_B | \mathcal{M}^{\gamma,\theta}(\hat{x}^R_A,\hat{x}^R_B) \neq \mathcal{M}^{\gamma,\theta}(x_A,x_B)\rbrace 
{P}_H \lbrace \mathcal{M}^{\gamma,\theta}(\hat{x}^R_A,\hat{x}^R_B) \neq \mathcal{M}^{\gamma,\theta}(x_A,x_B)\rbrace\\
\label{pe_ub}
& \leq \underbrace{{P}_H \lbrace E_A | \mathcal{M}^{\gamma,\theta}(\hat{x}^R_A,\hat{x}^R_B) = \mathcal{M}^{\gamma,\theta}(x_A,x_B)\rbrace}_{{P}_H^{A,BC}} 
+\underbrace{{P}_H \lbrace E_B | \mathcal{M}^{\gamma,\theta}(\hat{x}^R_A,\hat{x}^R_B) = \mathcal{M}^{\gamma,\theta}(x_A,x_B)\rbrace}_{{P}_H^{B,BC}} 
+\underbrace{{P}_H \lbrace \mathcal{M}^{\gamma,\theta}(\hat{x}^R_A,\hat{x}^R_B) \neq \mathcal{M}^{\gamma,\theta}(x_A,x_B)\rbrace}_{{P}_H^{MA}}.
\end{align}
\hrule
}\end{figure*}

The first and second terms in \eqref{pe_ub}, denoted as ${P}_H^{A,BC}$ and ${P}_H^{B,BC},$ are the probability of error events at node A and B respectively at the end of the BC phase, given that the relay decoded to the correct cluster during the MA phase  The third term in \eqref{pe_ub}, denoted as ${P}_H^{MA},$ is the probability that the relay decodes to the wrong cluster during the MA Phase.  

%We have ${P}_H^{A,BC}$ is the probability that $X_R=\mathcal{M}^{\gamma,\theta}(\hat{x}^R_A,\hat{x}^R_B)$ is wrongly decoded as $X'_R$ at node A and can be upper-bounded as,
%
%\begin{align}
%\nonumber
%{P}_H^{A,BC} &\leq \sum_{X'_R \neq X_R \in \mathcal{S}_R(\gamma, \theta)} Q[\vert  H'_A(X_R-X'_R)\vert]\\
%\label{pe_BC_ub}
%&\leq \sum_{X'_R \neq X_R \in \mathcal{S}_R(\gamma, \theta)} e^{-\frac{\vert H'_A(X_R-X'_R)\vert ^2}{2}}.
%\end{align}
%
%Let ${P}^{A,BC}=\mathbb{E}_H({P}_H^{A,BC}).$  To obtain an upper-bound on ${P}^{A,BC},$ \eqref{pe_BC_ub} needs to be averaged over different realizations of $H.$ Note that the signal set $\mathcal{S}_R(\gamma, \theta)$ chosen is dependent on the realization of $H_A$ and $H_B$ and calculating the expectation of the upper bound in \eqref{pe_BC_ub} is difficult.  

The choice of the signal set $\mathcal{S'}$ used at R during BC phase depends on the number of clusters in the clustering $\mathcal{C}^{\gamma, \theta}$ \cite{APT1},\cite{NVR}. Let $ \mathcal{C}_{max}$ denote the clustering which has the maximum number of clusters over all possible $\gamma e^{j \theta}$ and let $ \mathcal{S'}_{max}$ denote the signal set associated with the clustering $ \mathcal{C}_{max}.$ Let ${P}^{A,BC}=\mathbb{E}({P}_H^{A,BC})$ and ${P}^{B,BC}=E({P}_H^{B,BC}).$ The upper-bound on the average error probability during the BC phase calculated for the case when R always uses the signal set $ \mathcal{S'}_{max}$ irrespective of $\gamma e^{j \theta}$ will serve as an upper-bound on ${P}^{A,BC}$ and ${P}^{B,BC}$ as well. Hence, we have,
$${P}^{A,BC} = {P}^{B,BC} \leq e^{-K} \frac{1}{1+\frac{\text{SNR} \: d^2_{min}(\mathcal{S'}_{max})}{4}},$$ where $d_{min}(\mathcal{S'}_{max})$ is the minimum distance of the signal set $\mathcal{S'}_{max}.$

The upper-bound on ${P}^{A,BC}$ and ${P}^{B,BC}$ given above is not tight. Since the performance advantage due to adaptive network coding is captured only by the term ${P}_H^{MA}$ in \eqref{pe_ub}, in the rest of the paper we do not focus on the probabilities ${P}^{A,BC}$ and ${P}^{B,BC}.$  

The probability that the relay decodes to the wrong cluster during the MA Phase ${P}_H^{MA}$ can be upper-bounded as,

\begin{align}
\label{pe_MA_ub}
{P}_H^{MA} \leq \hspace{-.5 cm}\sum_{(x'_A,x'_B):\mathcal{M}^{\gamma,\theta}(x'_A,x'_B) \neq \mathcal{M}^{\gamma,\theta}(x_A,x_B)}\hspace{-1.5 cm}{P}_H \lbrace (\hat{x}^R_A,\hat{x}^R_B)=(x'_A,x'_B) \rbrace. 
\end{align} 

The probability ${P}_H \lbrace (\hat{x}^R_A,\hat{x}^R_B)=(x'_A,x'_B) \rbrace$ can be upper-bounded by the corresponding pair-wise error probability given by,

 {\footnotesize \begin{align*}P_H ^{MA} \lbrace(x_A,x_B) \rightarrow (x'_A,x'_B)\rbrace&=\\ & \hspace{-1 cm}Q\left[\frac{\sqrt{SNR}\vert H_A (x_A-x'_A)+H_B (x_B-x'_B) \vert}{\sqrt{2}}\right].\end{align*}} 

Hence, from \eqref{pe_MA_ub}, we get \eqref{pe_MA_ub1} (shown at the top of the next page), where $1_{\lbrace c \rbrace}$ is the indicator function which is one if the condition $c$ is satisfied and is zero if it is not satisfied. 

\begin{figure*}
{\scriptsize
\begin{align}
\nonumber
{P}_H^{MA} &\leq \hspace{-.1 cm}\sum_{(x'_A,x'_B):\mathcal{M}^{\gamma,\theta}(x'_A,x'_B) \neq \mathcal{M}^{\gamma,\theta}(x_A,x_B)}\hspace{-1.1 cm}Q\left[\frac{\sqrt{SNR}\vert H_A (x_A-x'_A)+H_B (x_B-x'_B) \vert}{\sqrt{2}}\right]\\ \label{pe_MA_ub1}
&=\sum_{(x'_A,x'_B):(x'_A,x'_B) \neq (x_A,x_B)} \hspace{-.1 cm}\left(\mathbf{1}_{\lbrace \mathcal{M}^{\gamma,\theta}(x'_A,x'_B) \neq \mathcal{M}^{\gamma,\theta}(x_A,x_B) \rbrace } Q\left[\frac{\sqrt{SNR}\vert H_A (x_A-x'_A)+H_B (x_B-x'_B) \vert}{\sqrt{2}}\right]\right),
\end{align}
\hrule
}\end{figure*}

Averaging with respect to the fade coefficients in \eqref{pe_MA_ub1}, we get \eqref{pe_MA_ub2} (shown at the top of the next page),  where $\mathcal{R}\lbrace (x_A,x_B),(x'_A,x'_B) \rbrace \subseteq \mathbb{C}^2$ is the region $$\lbrace (H_A,H_B):\mathcal{M}^{H_B/H_A}(x_A,x_B) \neq \mathcal{M}^{H_B/H_A}(x'_A,x'_B)\rbrace.$$

\begin{figure*}[t]
{\scriptsize
\begin{align}
\label{pe_MA_ub2}
&{P}^{MA} \triangleq \mathbb{E}({P}_H^{MA}) \leq \hspace{-.5 cm}\sum_{(x'_A,x'_B):(x'_A,x'_B) \neq (x_A,x_B)} \underbrace {\displaystyle{\int \int}_{\mathcal{R}\lbrace (x_A,x_B),(x'_A,x'_B) \rbrace}\hspace{-.3 cm}Q\left[\frac{\sqrt{SNR}\vert H_A (x_A-x'_A)+H_B (x_B-x'_B)\vert}{\sqrt{2}} \right] f(H_A) f(H_B) dH_A dH_B}_{P^{MA}\lbrace(x_A, x_B) \rightarrow (x'_A,x'_B) \rbrace}.\\
\hline
\label{pe_MA_ub3}
&P^{MA}_{FNC}\lbrace(x_A, x_B) \rightarrow (x'_A,x'_B) \rbrace={\int \int}_{\mathbb{C}^2}Q\left[\frac{\sqrt{SNR}\vert H_A (x_A-x'_A)+H_B (x_B-x'_B) \vert}{\sqrt{2}}\right] f(H_A) f(H_B) dH_A dH_B.
\end{align}
\hrule
}
\end{figure*}

%Let $\Delta x_A=x_A-x'_A$ and $\Delta x_B = x_B-x'_B.$  
 Let $P^{MA}\lbrace(x_A, x_B) \rightarrow (x'_A,x'_B) \rbrace$ denote the term inside the summation in \eqref{pe_MA_ub2}. In the following subsections upper bounds on $P^{MA}\lbrace(x_A, x_B) \rightarrow (x'_A,x'_B) \rbrace$ are obtained for the two-way relaying scenarios without and with adaptive network coding.

\subsection{Two-way Relaying without Adaptive Network Coding}

Consider the situation where R uses the same clustering $\mathcal{C}$ which does not remove any of the singular fade states, for all values of $\gamma e^{j\theta}.$ Since R uses the same clustering for all $\gamma e^{j\theta},$ the region $\mathcal{R}\lbrace (x_A,x_B),(x'_A,x'_B)\rbrace$ can be either null set or the entire $\mathbb{C}^2$ plane. If $(x_A,x_B)$ and $(x'_A,x'_B)$ are placed in the same cluster by the clustering $\mathcal{C},$ then the region is the null set and $P^{MA}\lbrace(x_A, x_B) \rightarrow (x'_A,x'_B) \rbrace=0.$ But since $\mathcal{C}$ does not remove any of the singular fade states, for every singular fade state $-\frac{\Delta x_A}{\Delta x_B},\Delta x_A,\Delta x_B \in \Delta \mathcal{S},$ there exists at least one pair of two-tuples $\lbrace(x_A,x_B),(x'_A,x'_B)\rbrace$ which satisfies $\Delta x_A=x_A-x'_A$ and $\Delta x_B = x_B-x'_B$ and for which $(x_A,x_B)$ and $(x'_A,x'_B)$ are not placed in the same cluster.  For such pairs $(x_A,x_B)$ and $(x'_A,x'_B),$ $\mathcal{R}\lbrace (x_A,x_B),(x'_A,x'_B)\rbrace$ is the entire $\mathbb{C}^2$ plane. For this case, the probability $P^{MA}\lbrace(x_A, x_B) \rightarrow (x'_A,x'_B) \rbrace$ defined in \eqref{pe_MA_ub} is given in \eqref{pe_MA_ub3} (shown at the top of the next page). In \eqref{pe_MA_ub3}, the suffix $FNC$ indicates that fixed network coding is used at R, irrespective of the channel conditions. Also, in \eqref{pe_MA_ub3}, $f(H_A)$ and $f(H_B)$ denote the probability density functions of the random variables $H_A$ and $H_B$ respectively. Substituting Rician probability density functions for $f(H_A)$ and $f(H_B),$ at high SNR, the integral given in \eqref{pe_MA_ub3} can be upper bounded as,

{\footnotesize
\begin{align}
\label{pe_MA_ub4}
&P^{MA}_{FNC}\lbrace(x_A, x_B) \rightarrow (x'_A,x'_B) \rbrace < \frac{e^{-K \left(\frac{\vert \Delta x_A + \Delta x_B \vert ^2}{\vert \Delta x_A \vert ^2 + \vert \Delta x_B \vert ^2}\right)}}{\left(1+\frac{\text{SNR}\left(\vert \Delta x_A \vert ^2 + \vert \Delta x_B \vert ^2\right)}{4}\right)}.
\end{align}
}

From \eqref{pe_MA_ub4}, it can be seen that $P^{MA}_{FNC}\lbrace(x_A, x_B) \rightarrow (x'_A,x'_B)\rbrace$ decreases as $\text{SNR}^{-1}$ at high SNR.

The probabilities $P^{A,{BC}}$ and $P^{B,{BC}}$ are proportional to $e^{-K}$ at high SNR. If $\Delta x_B$ belongs to $\Delta \mathcal{S},$ $-\Delta x_B$ also belongs to $\Delta \mathcal{S}.$ For $\Delta x_A \neq 0$ and $\Delta x_B \neq 0,$ one of the two ${\vert \Delta x_A + \Delta x_B \vert ^2 }$ and ${\vert \Delta x_A - \Delta x_B \vert ^2 },$ has to be less than $\vert \Delta x_A \vert ^2 +\vert \Delta x_B \vert ^2.$ Hence, from \eqref{pe_MA_ub4}, $P^{MA}_{FNC}\lbrace(x_A, x_B) \rightarrow (x'_A,x'_B) \rbrace$  is proportional to $e^{-\kappa K}$ where $\kappa <1,$ for some $x_A \neq x'_A$ and $x_B \neq x'_B.$ 
 When the Rician factor $K$ increases, the contribution of the error during the BC phase to the overall average SER decreases and for large values of $K,$ the contribution to the overall average SER comes totally from the error during the MA phase. The reason for this is that $P^A_{BC}$ and $P^B_{BC}$ (proportional to $e^{-K}$) decrease faster with $K$ than those terms $P^{MA}\lbrace(x_A, x_B) \rightarrow (x'_A,x'_B) \rbrace$ which are proportional to $e^{-\kappa K}$ where $\kappa <1.$ For this reason, adaptive network coding, which improves the performance during the MA phase, provides more gain in Rician fading scenario than in a Rayleigh fading scenario, consistent with the simulation results in \cite{APT1} and \cite{VNR}. The exact reason why adaptive network coding improves the performance during the MA phase is described in the next subsection.
\subsection{Two-way Relaying with Adaptive Network Coding}
Consider the adaptive network coding schemes proposed in \cite{APT1} and \cite{NVR}. In both the schemes all the singular fade states  are removed by a proper choice of the clustering. 

Consider error events $(x_A, x_B) \rightarrow (x'_A,x'_B)$ at R for which $x_A=x'_A$ or $x_B=x'_B.$ For such error events, the region $\mathcal{R}\lbrace (x_A,x_B),(x'_A,x'_B)\rbrace$ is the entire $\mathbb{C}^2$ since the pairs  $(x_A, x_B)$ and $(x'_A, x_B)$ (and also the pairs $(x_A, x_B)$ and $(x_A, x'_B)$)  cannot be placed in the same cluster without violating the exclusive law. At high SNR, the probabilities $P^{MA}\lbrace(x_A, x_B) \rightarrow (x_A,x'_B) \rbrace$ and $P^{MA}\lbrace(x_A, x_B) \rightarrow (x'_A,x_B) \rbrace$ defined in \eqref{pe_MA_ub2} can be upper-bounded as,

\begin{align}
\label{non_rem_1}
 &P^{MA}_{ANC}\lbrace(x_A, x_B) \rightarrow (x_A,x'_B) \rbrace < \frac{e^{-K}}{1+\frac{\text{SNR} \: \vert \Delta x_B \vert ^2}{4}}, \\
\label{non_rem_2} 
 &P^{MA}_{ANC}\lbrace(x_A, x_B) \rightarrow (x'_A,x_B) \rbrace < \frac{e^{-K}}{1+\frac{\text{SNR} \: \vert \Delta x_A \vert ^2}{4}}.
\end{align} 
The suffix $ANC$ in the above two equations indicates that adaptive network coding is used at R.

\begin{figure}[htbp]
\vspace{-1 cm}
\centering
\includegraphics[totalheight=3.75in,width=4in]{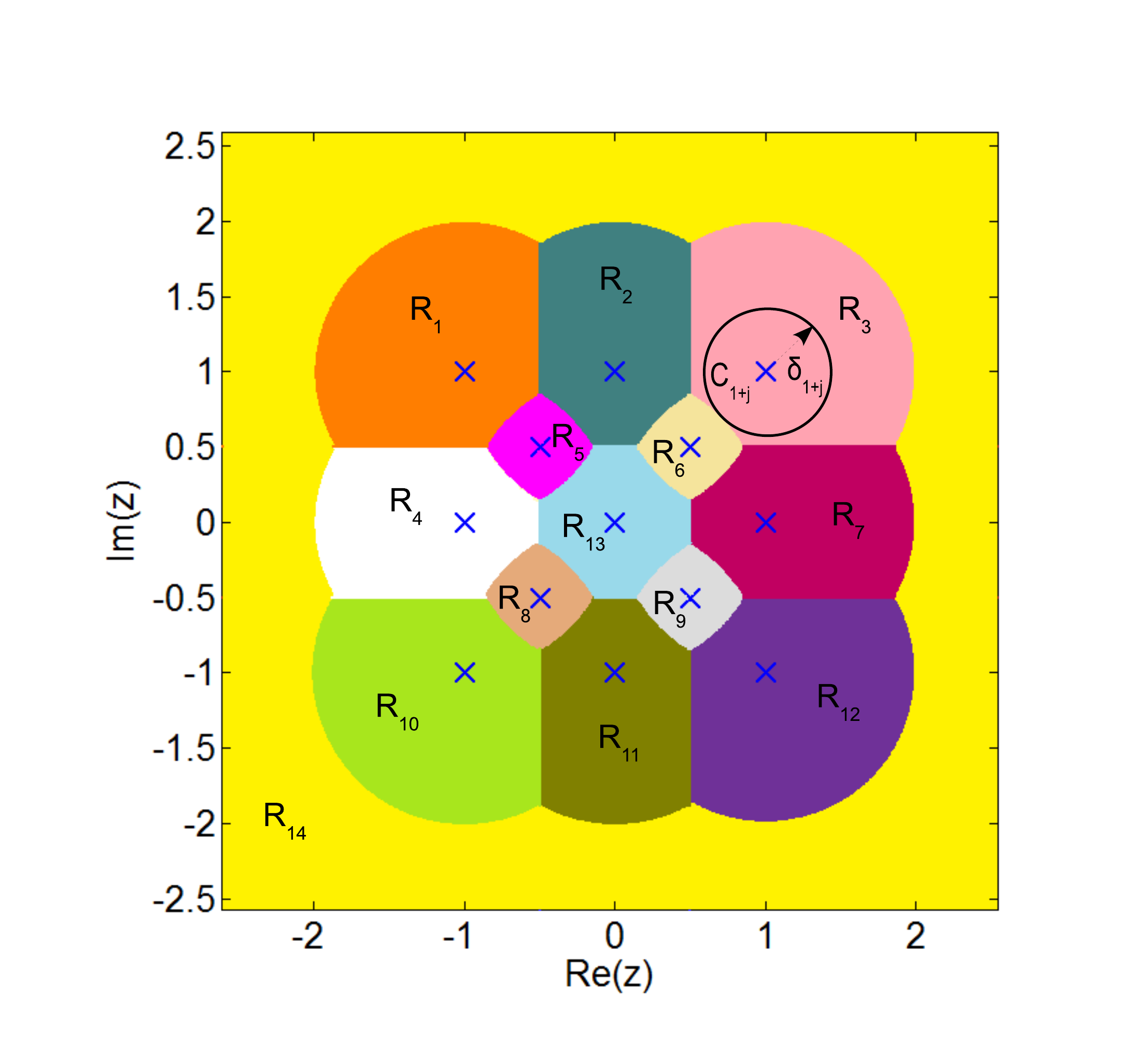}
\vspace{-1 cm}
\caption{The Diagram showing the fade state quantization for 4-PSK signal set}
\label{4psk_partition}
\end{figure}

Let $s=-\frac{\Delta x_A}{\Delta x_B},\Delta x_A,\Delta x_B \neq 0 \in \Delta \mathcal{S}$ denote a singular fade state. For the adaptive network coding schemes proposed in \cite{APT1} and \cite{VNR}, the complex fade state ($\gamma e^{j \theta}$) plane can be quantized into different regions depending on the clustering used at R. In the neighbourhood of every singular fade state, an associated region ${\Gamma}_s$ exists in which a clustering which removes that singular fade state is used at R. For example, for the case when 4-PSK signal set is used at A and B, the quantization of the complex fade state plane is as shown in Fig. \ref{4psk_partition}, along with the regions $\text{R}_1-\text{R}_{12}$ associated with the 12 singular fade states \cite{VNR}. The regions $\text{R}_{13}$ and $\text{R}_{14}$ in Fig. \ref{4psk_partition} are the clustering independent regions in which choice of the clustering does not matter and any clustering satisfying the exclusive law gives the same performance. For details, see \cite{VNR}.

\begin{note}
  The quantization of the complex fade state plane is the same for 4-PSK signal set for the Nearest Neighbour Clustering algorithm proposed in \cite{NVR} and for the scheme proposed in \cite{APT1}, while it need not be the same for other signal sets. Nevertheless, for both the schemes, there exists a region in the neighbourhood of every singular fade state in which a clustering which removes that singular fade state will be used. 
  \end{note}
  
  Let $C_s$ denote the circle with the largest radius $\delta_s$ enclosed in the region $\Gamma_s,$ with center at the singular fade state $s.$ For example, for 4-PSK signal set, $s=1+j$ is a singular fade state and the circle $C_{1+j}$ of radius $\delta_{1+j}$ enclosed in the region $\Gamma_{1+j}$ (the region $\text{R}_3$) is as shown in Fig. \ref{4psk_partition}.

 Let $\mathcal{E}(C_s)$ denote the region exterior to the circle $C_s$ in the complex plane. If $\gamma e^{j \theta}$ lies inside the circle $C_s,$ $\mathcal{M}^{\gamma,\theta}(x_A,x_B)= \mathcal{M}^{\gamma,\theta}(x'_A,x'_B).$ Hence, $\mathcal{R}\lbrace (x_A,x_B),(x'_A,x'_B) \rbrace \subset \lbrace (H_A,H_B) : \frac{H_B}{H_A} \in \mathcal{E}(C_s) \rbrace = \lbrace (H_A,H_B) : \vert \frac{H_B}{H_A} +\frac{\Delta x_A}{\Delta x_B} \vert \geq \delta_s \rbrace.$ Hence, for the adaptive network coding schemes, the probability $P^{MA}\lbrace(x_A, x_B) \rightarrow (x'_A,x'_B) \rbrace$ defined in \eqref{pe_MA_ub2} can be upper-bounded as given in \eqref{pe_MA_adaptive_ub} (the suffix $ANC$ in \eqref{pe_MA_adaptive_ub} indicates that adaptive network coding is used at R). 
\begin{figure*}
{\scriptsize
\begin{align}
\label{pe_MA_adaptive_ub}
&P^{MA}_{ANC}\lbrace(x_A, x_B) \rightarrow (x'_A,x'_B) \rbrace \leq {\int \int}_{\left\lbrace \left(H_A,H_B \right) : \left \vert \frac{H_B}{H_A} +\frac{\Delta x_A}{\Delta x_B} \right \vert \geq \delta_s \right \rbrace.}Q\left[\frac{\sqrt{SNR}\vert H_A (x_A-x'_A)+H_B (x_B-x'_B) \vert}{\sqrt{2}}\right] f(H_A) f(H_B) dH_A dH_B.
\end{align}
\hrule
}
\end{figure*}
Let $\Delta x_A =x_A-x'_A$ and $\Delta x_B = x_B-x'_B.$ At high SNR, $P^{MA}_{ANC}\lbrace(x_A, x_B) \rightarrow (x'_A,x'_B) \rbrace$ can be upper-bounded as stated in the following theorem. Note that the bound is valid only at high SNR and is obtained by upper-bounding the term on the right hand side of \eqref{pe_MA_adaptive_ub}.  

\begin{theorem}
For $x_A \neq x'_A$ and $x_B \neq x'_B,$ at high SNR $P^{MA}_{ANC}\lbrace(x_A, x_B) \rightarrow (x'_A,x'_B) \rbrace$ can be upper-bounded as given in \eqref{pe_MA_adaptive_ub2} (shown at the top of the next page). 
{\begin{figure*}
\scriptsize
\begin{align}
\label{pe_MA_adaptive_ub2}
P^{MA}_{ANC}\lbrace(x_A, x_B) \rightarrow (x'_A,x'_B) \rbrace \leq \frac{\exp \left \lbrace -2K +\frac{K(K+1)\left \vert 1-\frac{\Delta x_A }{ \Delta x_B} \right\vert^2}{\left[K+1+\text{SNR}\frac{\vert \Delta x_B \vert ^2}{4} \right]\delta_s^2 +(K+1)\left(1+\frac{\vert \Delta x_A \vert ^2}{\vert \Delta x_B \vert^2}\right)}\right \rbrace} {\left(K+1+\text{SNR}\frac{\vert \Delta x_B \vert^2}{4}\right)\left(\left[K+1+\text{SNR}\frac{\vert \Delta x_B \vert ^2}{4} \right]\delta_s^2 +(K+1)\left(1+\frac{\vert \Delta x_A \vert ^2}{\vert \Delta x_B \vert^2}\right)\right)}
\end{align}
\hrule
\end{figure*}
}
\begin{proof}
See Appendix.
\end{proof}
\end{theorem}

From Theorem 1, it can be seen that for $x_A \neq x'_A$ and $x_B \neq x'_B,$ at high SNR, $P^{MA}_{ANC}\lbrace(x_A, x_B) \rightarrow (x'_A,x'_B) \rbrace$ decreases as $\text{SNR}^{-2}.$ But the overall diversity order of the end-to-end SER will be one, since $P^{MA}_{ANC}\lbrace(x_A, x_B) \rightarrow (x_A,x'_B) \rbrace$ and $P^{MA}_{ANC}\lbrace(x_A, x_B) \rightarrow (x'_A,x'_B) \rbrace$ given in \eqref{non_rem_1} and \eqref{non_rem_2} as well as the probabilities $P^{A,BC}$ and $P^{B,BC}$ decrease as $\text{SNR}^{-1}$ at high SNR.

Even though adaptive network coding does not provide any diversity advantage, it provides coding gain advantage over the case when adaptive network coding is not used. For the case when adaptive network coding is not used, from \eqref{pe_MA_ub4}, it can be seen that  $P^{MA}_{FNC}\lbrace(x_A, x_B) \rightarrow (x'_A,x'_B) \rbrace$ decreases as $\text{SNR}^{-1}$ at high SNR, for all pairs $(x_A, x_B)$ and $(x'_A,x'_B).$ In contrast for the adaptive network coding scheme, only those  probabilities $P^{MA}_{ANC}\lbrace(x_A, x_B) \rightarrow (x'_A,x'_B) \rbrace$ for which $x_A=x'_A$ or $x_B=x'_B$ decrease as $\text{SNR}^{-1}$ and the rest decrease as $\text{SNR}^{-2}.$ In other words, for the adaptive network coding scheme, at high SNR, the pair-wise error events associated with the non-removable singular fade states 0 and $\infty$ decrease as $\text{SNR}^{-1}$ and the pair-wise error events associated with the removable singular fade states decrease as $\text{SNR}^{-2}.$ The adaptive network coding scheme provides coding gain advantage over the case when adaptive network coding is not used, by making  the probability of the error events $(x_A, x_B) \rightarrow (x'_A,x'_B)$ for which $x_A \neq x'_A$ and $x_B \neq x'_B$ decrease as $\text{SNR}^{-2}$ instead of $\text{SNR}^{-1}.$

%Also, as the Rician factor increases, for the case when adaptive network coding is not used, the error probabilities $P^{MA}\lbrace(x_A, x_B) \rightarrow (x_A,x'_B) \rbrace$ and $P^{MA}\lbrace(x_A, x_B) \rightarrow (x'_A,x'_B) \rbrace$ are proportional to $e^{-K}$ and they do not contribute much to the overall error probability. Those error probabilities $P^{MA}\lbrace(x_A, x_B) \rightarrow (x'_A,x'_B) \rbrace$ for which $x_A \neq x'_A$ and $x_B \neq x'_B$ contribute dominantly to the overall error probability. On the other hand, for the adaptive network coding scheme, at high SNR, only those error probabilities $P^{MA}\lbrace(x_A, x_B) \rightarrow (x_A,x'_B) \rbrace$ and $P^{MA}\lbrace(x_A, x_B) \rightarrow (x'_A,x'_B) \rbrace$ contribute to the overall error probability. To sum, adaptive network coding provides performance improvement over the case when adaptive network coding is not employed, by totally removing the effect of all the dominant error probability terms.

From \eqref{pe_MA_ub4}, for the case when adaptive network coding is not used, it follows that $P^{MA}_{FNC}\lbrace(x_A, x_B) \rightarrow (x'_A,x'_B) \rbrace$ is proportional to $e^{-K \frac{\vert1 -s \vert ^2}{1+\vert s \vert ^2}}.$ This suggests that removing certain singular fade states assumes more significance than removing the others and is discussed in the following section.

\section{Partial Removal of Singular Fade States} 
%In the adaptive network coding schemes proposed in \cite{APT1} and \cite{NVR}, for a given realization of the channel fade state, the choice of the network coding map is indicated to the end nodes using overhead bits. In \cite{APT1} an algorithm called the Closest Neighbour Clustering (CNC) algorithm was proposed which runs for a given $\gamma e^{j\theta}.$ The total number of network coding maps is known only after the CNC algorithm is run for all possible realizations of $\gamma e^{j \theta}$ which is uncountably infinite and is not known beforehand. The alternative procedure suggested in \cite{NVR} is based on the removal of all the singular fade states. Since the number of singular fade states increases with increase in the cardinality of the signal set, the number of overhead bits required also increases. For example, when $2^{\lambda}$-PSK signal set is used at A and B, the number of singular fade states is $(\frac{2^{2\lambda}}{4}-\frac{2^{\lambda}}{2}+1)2^{\lambda}$ and the number of overhead bits is approximately $3 \lambda$ for PSK constellations of large size.

 In this section, it is shown that only some of the removable singular fade states contribute dominantly to the overall SER and only such singular fade states can be removed without a significant degradation in performance. Note that whether a singular fade state contributes significantly to the error probability depends on the Rician factor as well. For a Rayleigh fading scenario (Rician factor K=0), contributions of all the singular fade states to the overall SER are significant. 
 
Since $P^{MA}_{FNC}\lbrace(x_A, x_B) \rightarrow (x'_A,x'_B) \rbrace$ is proportional to $e^{-K \frac{\vert1 -s \vert ^2}{1+\vert s \vert ^2}},$ the factor $f(s)=\frac{\vert1 -s \vert ^2}{1+\vert s \vert ^2},$ referred to as the \textit{dominance factor} of the singular fade state $s,$ determines whether the contribution from a singular fade state $s$ is significant or not. The lesser the value of the dominance factor, the more the contribution of the singular fade state $s$ to the SER. Since $f(s)=0$ if and only if $s=1,$ \textit{the removal of the singular fade state $s=1$ assumes greatest significance}. Note that for the case when A and  B use the same signal set, $s=1$ will always be a singular fade state. Let $\angle{s}$ denote the phase angle of $s.$ We have, $f(s)=1-\frac{2 \vert s \vert \cos \left(\angle{s} \right)}{1+\vert s \vert^2}.$  

%For a singular fade state $s,$ there can exists many pairs $(\Delta x_A, \Delta x_B)$ for which $s=-\frac{\Delta x_A}{\Delta x_B}.$ Let $\mathcal{L}_s$ denote all such pairs. Let $g(s)$ denote $\displaystyle \min_{(\Delta x_A, \Delta x_B) \in \mathcal{L}_s} (\vert \Delta x_A \vert^2 +\vert \Delta x_B \vert^2).$ The relay can choose to remove only those singular fade states for which $\frac{e^{-K f(s)} g(1)}{g(s)}>\tau,$ where $\tau$ is a fixed number less than 1 (say $0.1$).

Among those singular fade states which have the same absolute value $\vert s \vert,$ those which have a lesser value of $\vert \angle{s} \vert$ ($\vert \angle{s} \vert$ ranges from 0 to $\pi$) have a lesser dominance factor and hence contribute more towards the overall SER. For sufficiently large Rician factors, among those singular fade states which lie on the same circle, R can choose to remove only those singular fade states which have the least absolute value of the phase angle. 

\begin{example}
For the case when 4-PSK signal set is used at the nodes A and B, the twelve singular  fade states are as shown in Fig. \ref{4psk_sing}. 
For 4-PSK signal set, the dominance factor $f(s)$ for the 12 singular fade states are given by,

{\centering
$$\begin{array}{|c|c|}
\hline s & f(s) \\
%\hline 0,\infty & 1 \\
\hline 1& 0  \\
\hline j,-1,-j& 1\\
%\hline -1& 1\\
%\hline -j& 1\\
\hline 0.5+0.5j,0.5-0.5j,1+j,1-j & {1}/{3} \\
\hline -0.5+0.5j,-0.5-0.5j,-1+j,-1-j & {5}/{3}\\
%\hline -0.5-0.5j& {5}/{3}\\
%\hline 0.5-0.5j &{1}/{3}\\
%\hline 1+j& {1}/{3}\\
%\hline -1+j&{5}/{3}\\
%\hline -1-j& {5}/{3}\\
%\hline 1-j&{1}/{3} \\
\hline
\end{array}$$
}
Among all the singular fade states, the singular fade state $s=1$ is the most dominant one. Among those singular fade states which lie on the circle with radius $\frac{1}{\sqrt{2}},$ the singular fade states $0.5+0.5j$ and $0.5-0.5j$ are the dominant ones. Similarly, among those singular fade states which lie on the circle with radius $\sqrt{2},$ the singular fade states $1+j$ and $1-j$ are the dominant ones. The relay can choose to remove only the dominant singular fade states on each circle, which are the circled ones shown in Fig. \ref{4psk_sing}.
\begin{figure}[htbp]
\vspace{-.9 cm}
\centering
\includegraphics[totalheight=3.75in,width=4in]{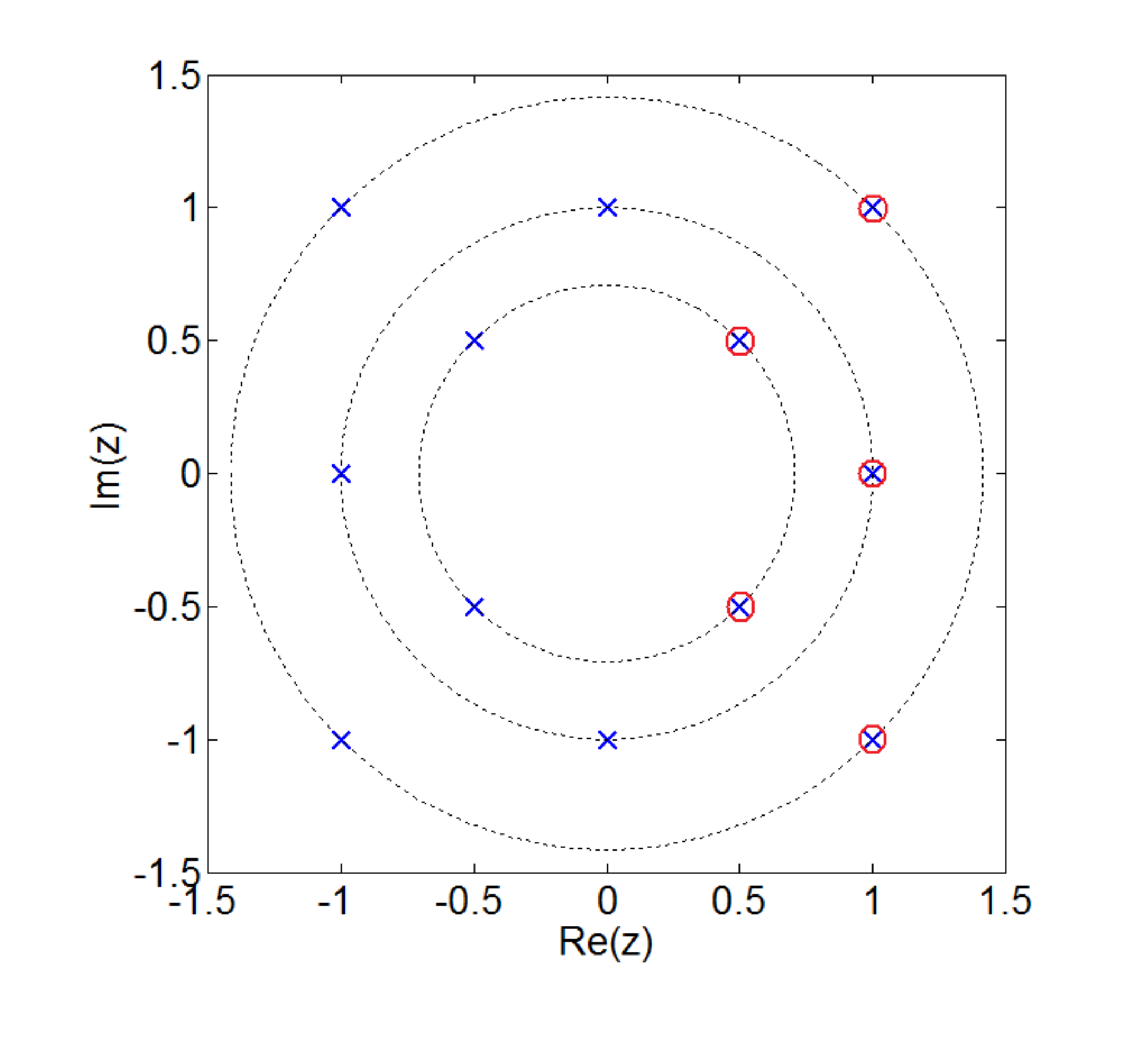}
\vspace{-1 cm}
\caption{The Diagram showing the dominant singular fade states for 4-PSK signal set}
\label{4psk_sing}
\end{figure}
\end{example}

\section{Simulation Results}
\begin{figure}[htbp]
\vspace{-.5 cm}
\includegraphics[totalheight=3.4in,width=3.8in]{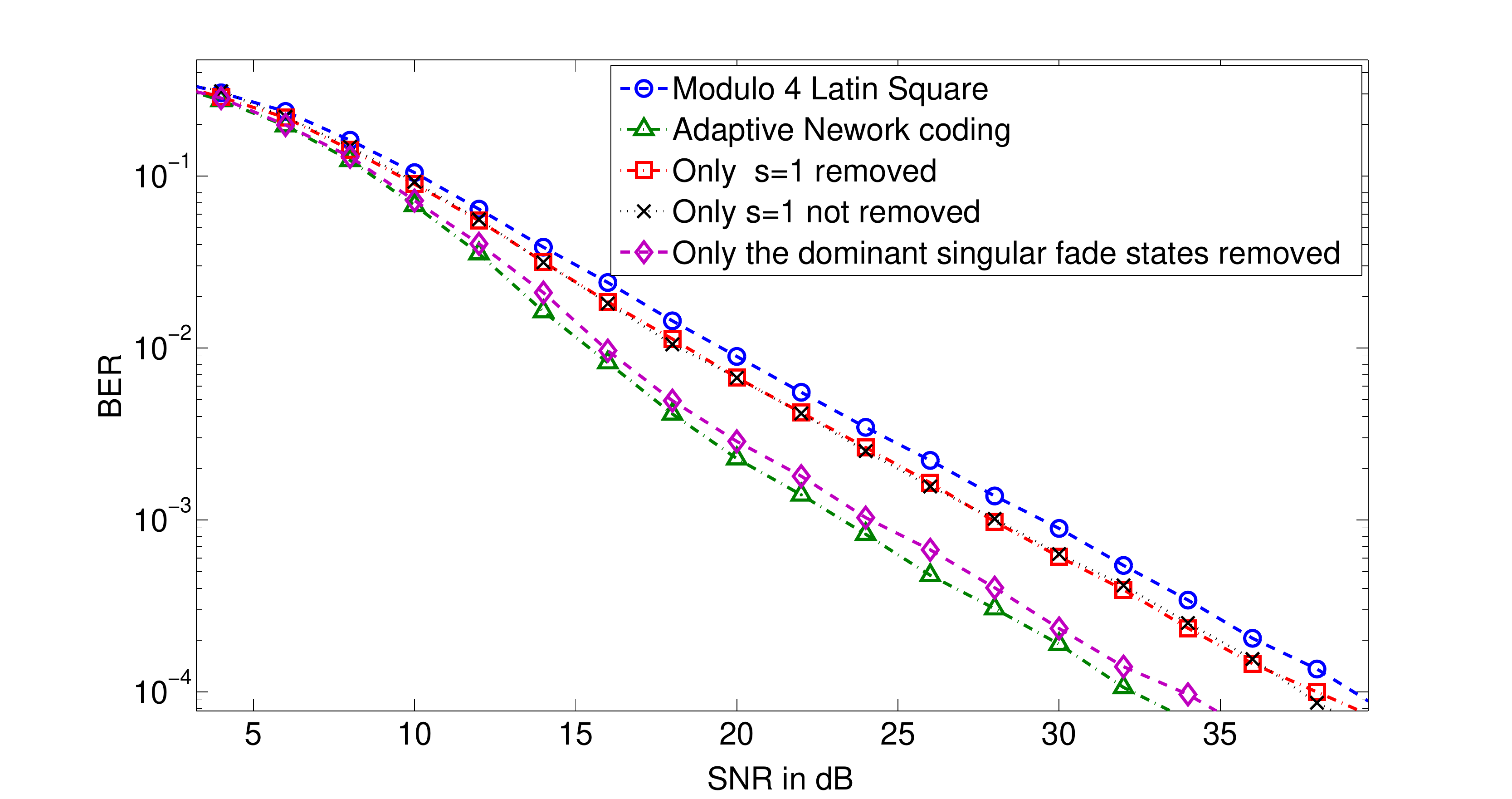}
\vspace{-.5 cm}
\caption{SNR vs BER plots for the different schemes for Rician factor $K=4$}
\label{simulation_plot}
\end{figure}

The simulation results presented are for the case when the end nodes use 4-PSK signal set. The fading coefficient $H_A,$ $H_B,$ $H'_A,$ $H'_B$ are Rician distributed with a Rician factor $K=4$ and unit variance. For comparison, we consider the case when R uses the Modulo-4 Latin Square shown in Fig. \ref{modulo_4_Latin} irrespective of the channel condition (every entry of the Modulo-4 Latin Square is the modulo 4 addition of the row index and the column index). Note that the Modulo-4 Latin Square does not remove any of the 12 singular fade states. Fig. \ref{simulation_plot} shows the SNR vs BER plots for the different cases. From Fig. \ref{simulation_plot}, it can be seen that the diversity order for all the cases considered is one. Also, it can be seen that at high SNR, the adaptive network coding scheme based on the removal of all the singular fade states using Latin Squares proposed in \cite{NVR} provides nearly 8 dB gain over the case when Modulo-4 Latin Square is used irrespective of channel conditions. For details regarding the Latin Squares which remove the singular fade states, see \cite{NVR}. 

\begin{note}
For 4-PSK signal set, the adaptive network coding scheme based on the removal of singular fade states proposed in \cite{NVR} and the one based on the Nearest Neighbour Clustering algorithm proposed in \cite{APT1} turn out to be the same.
\end{note}

\begin{figure}[h]
\centering
\begin{tabular}{|c|c|c|c|c|}
\hline &0 & 1 & 2 & 3\\ 
\hline 0&0 & 1 & 2 & 3\\ 
\hline 1&1 & 2 & 3 & 0\\ 
\hline 2&2 & 3 & 0 & 1\\ 
\hline 3& 3 & 0 & 1 & 2\\ 
\hline 
\end{tabular}
\caption{The Modulo-4 Latin Square}
\label{modulo_4_Latin}
\end{figure}

Fig. \ref{simulation_plot} also shows the SNR vs BER plots for the case when only the singular fade state $s=1$ is removed and for the case when all the singular fade states other than $s=1$ are removed. Note that the bit-wise XOR network code removes the singular fade state $s=1$ \cite{NVR}. It can be seen that the SNR vs BER performance for  both these cases are nearly the same. This means that the performance improvement provided by removing all the 11 singular fades states other than $s=1,$ is equal to the performance improvement provided by the removal of $s=1$ alone. This confirms the assertion made earlier in Section III that the removal of the singular fade state $s=1$ is of greatest significance. Fig. \ref{simulation_plot} shows the SNR vs BER plot for the case when only the five dominant singular fade states (the circled ones shown in Fig. \ref{4psk_sing}) are removed. It can be seen from Fig. \ref{simulation_plot} that at high SNR, removing only the five dominant singular fade states results in nearly 7 dB performance improvement over the case when Modulo-4 Latin Square is used irrespective of channel conditions. In other words, at high SNR, the degradation in performance that results because of removing not all but only the dominant singular fade states is less than 1 dB.   

\section{Discussion}
An upper bound on the average end-to-end symbol error probability was obtained for the two-way relaying scenarios with and without adaptive network coding. From the analysis, the reason why adaptive network coding schemes provide performance improvement becomes clear. Also, it was shown that in a Rician fading scenario, some of the singular fade states contribute more to the average symbol error probability. Simulation results show that removing only such dominant singular fade states results in almost the same performance as that of the case when all the singular fade states were removed. 
\section*{Acknowledgement}
This work was supported  partly by the DRDO-IISc program on Advanced Research in Mathematical Engineering through a research grant as well as the INAE Chair Professorship grant to B.~S.~Rajan.
\begin{figure*}[t]
{\scriptsize
\begin{align}
\label{Int_HA_1}
I(H_A) &= { \int}_{\left\lbrace H_B  : \left \vert {H_B} +\frac{H_A \Delta x_A}{\Delta x_B} \right \vert \geq \vert H_A \vert \delta_s \right \rbrace.}Q\left[\frac{\sqrt{SNR}\vert H_A (x_A-x'_A)+H_B (x_B-x'_B) \vert}{\sqrt{2}}\right]  f(H_B)  dH_B\\
\label{Int_HA_2}
&\leq \frac{K+1}{\pi}{ \int}_{\left\lbrace H_B  : \left \vert {H_B} +\frac{H_A \Delta x_A}{\Delta x_B} \right \vert \geq \vert H_A \vert \delta_s \right \rbrace} \exp\left\lbrace{- \frac{SNR\vert H_A \Delta x_A + H_B \Delta x_B \vert ^2}{4}}\right\rbrace \exp\left\lbrace{- (K+1)\left  \vert H_B-\sqrt{\frac{K}{K+1}} \right \vert ^2}\right\rbrace d H_B\\
\label{Int_HA_3}
&= \frac{K+1}{\pi}{ \int}_{\left\lbrace H'_B  : \left \vert {H'_B}  \right \vert \geq \vert H_A \vert \delta_s \right \rbrace} \exp\left \lbrace{- \left(K+1+\frac{SNR\vert \Delta x_B  \vert ^2}{4} \right) \left \vert H'_B - \frac{ (K+1)c}{\left[K+1+SNR\frac{\vert \Delta x_B  \vert ^2}{4}\right]}\right \vert ^2} \right\rbrace \exp\left\lbrace{- \frac{\vert c  \vert ^2 (K+1) \frac{SNR\vert \Delta x_B \vert ^2}{4}}{K+1+\frac{SNR\vert \Delta x_B \vert^2}{4}}}\right\rbrace d H'_B \\
\nonumber
& \leq \frac{1}{\left[ 1+K+\frac{SNR\vert \Delta x_B \vert ^2 }{4} \right]}  \frac{ \delta_s \vert H_A \vert }{\left[\delta_s \vert H_A \vert-\frac{(K+1) \vert{c}\vert}{K+1+SNR\frac{\vert \Delta x_B  \vert ^2}{4}}\right]} \exp\left \lbrace{- \left [ 1+K+ \frac{SNR\vert \Delta x_B \vert ^2 }{4} \right] \left ( \delta_s \vert H_A \vert -\frac{(K+1) \vert{c}\vert}{K+1+\frac{SNR\vert \Delta x_B  \vert ^2}{4}}\right)^2}\right \rbrace \\
\label{Int_HA_4}
&\hspace{12 cm}\exp\left\lbrace{- \frac{\vert c  \vert ^2 (K+1) \frac{SNR\vert \Delta x_B \vert ^2}{4}}{K+1+\frac{SNR\vert \Delta x_B \vert^2}{4}}}\right\rbrace \\
 \label{Int_HA_5}
& \approx \frac{1}{\left[ 1+K+\frac{SNR\vert \Delta x_B \vert ^2 }{4} \right]}   \exp\left \lbrace{- \left [ 1+K+ \frac{SNR\vert \Delta x_B \vert ^2 }{4} \right]  \delta_s^2 \vert H_A\vert ^2 } \right \rbrace \exp\left\lbrace{- \vert c  \vert ^2 (K+1)}\right\rbrace
%{e^{-(K+1)\frac{\vert \Delta x_B \vert ^2 }{4} \left \vert \frac{H_A \Delta x_A}{\Delta x_B} + \sqrt{\frac{K}{K+1}} \right \vert ^2 }
\end{align}
}\hrule
\end{figure*}
 %%%%%%%%%%%%%%%%%%%%%%%%%%%%%%%%%%%%%%%%%%%%%%%%%%%%%%%%%%%%%%%%%%%%%%%%%%%%%%%%%%%%%%%

\begin{appendix}
\begin{figure*}
{\scriptsize
\begin{align}
\nonumber
P_{ANC}^{MA}{(x_A,x_B)\rightarrow  (x'_A,x'_B)} &\leq \frac{1}{K+1+\frac{\text{SNR}\vert \Delta x_B \vert^2}{4}} \int_{H_A \in \mathbb{C}} \exp \left \lbrace - \left[K+1+\frac{\text{SNR}\vert \Delta x_B \vert ^2}{4} \right]\delta_s^2 \vert H_A \vert ^2\right \rbrace \exp\left \lbrace -(K+1)\left\vert H_A\frac{\Delta x_A}{\Delta x_B}+\sqrt{\frac{K}{K+1}}\right\vert^2 \right\rbrace \\
\label{proof_start2}
&\hspace{8 cm}\exp \left \lbrace -(K+1)\left \vert H_A-\sqrt{\frac{K}{K+1}} \right\vert^2\right\rbrace d H_A\\
\nonumber
&=\frac{1}{K+1+\frac{\text{SNR}\vert \Delta x_B \vert^2}{4}} \int_{H_A \in \mathbb{C}: \vert H_A \vert \geq 0} \exp \left \lbrace -2K +\frac{K(K+1)\vert 1+s \vert^2 }{\left[K+1+\frac{\text{SNR}\vert \Delta x_B \vert ^2}{4} \right]\delta_s^2 +(K+1)(1+\vert s \vert ^2)}\right \rbrace\\
\label{proof_start3}
& \hspace{-.5 cm} \exp \left \lbrace - \left(\frac{\left[K+1+\frac{\text{SNR}\vert \Delta x_B \vert ^2}{4} \right]\delta_s^2 + (K+1)(1+\vert s \vert ^2)}{\vert 1+s \vert^2}\right) \left \vert H_A (1+s)-\frac{\sqrt{K(K+1)}\vert 1+s \vert^2}{\left[K+1+\frac{\text{SNR}\vert \Delta x_B \vert ^2}{4} \right]+(K+1)(1+\vert s \vert^2)}\right \vert ^2\right \rbrace dH_A\\
\nonumber
&=\frac{\exp \left \lbrace -2K +\frac{K(K+1)\vert 1+ s \vert ^2}{\left[K+1+\frac{\text{SNR}\vert \Delta x_B \vert ^2}{4} \right]\delta_s^2 +(K+1)(1+\vert s \vert ^2)}\right \rbrace} {\left(K+1+\frac{\text{SNR}\vert \Delta x_B \vert^2}{4}\right)\left(\left[K+1+\frac{\text{SNR}\vert \Delta x_B \vert ^2}{4} \right]\delta_s^2 +(K+1)(1+\vert s \vert ^2)\right)} \\
\label{proof_start4}
&\hspace{1 cm}Q_1 \left[\sqrt{2\left(\left[K+1+\frac{\text{SNR}\vert \Delta x_B \vert ^2}{4} \right]\delta_s^2 +(K+1)(1+\vert s \vert ^2)\right)}\frac{\sqrt{K(K+1)}\vert 1+s \vert}{\left[K+1+\frac{\text{SNR}\vert \Delta x_B \vert ^2}{4} \right]+(K+1)(1+\vert s \vert^2)},0 \right]\\
\label{proof_start5}
&=\frac{\exp \left \lbrace -2K +\frac{K(K+1)\vert 1+ s \vert ^2}{\left[K+1+\frac{\text{SNR}\vert \Delta x_B \vert ^2}{4} \right]\delta_s^2 +(K+1)(1+\vert s \vert ^2)}\right \rbrace} {\left(K+1+\frac{\text{SNR}\vert \Delta x_B \vert^2}{4}\right)\left(\left[K+1+\frac{\text{SNR}\vert \Delta x_B \vert ^2}{4} \right]\delta_s^2 +(K+1)(1+\vert s \vert ^2)\right)}.
\end{align}
\hrule
}\end{figure*}
Before we prove Theorem 1, we prove the following lemma.

\begin{lemma}
For $c_0, r \in \mathbb{R}^{+}$ and $h_c \in \mathbb{C},$ the integral 
\begin{align}
\nonumber
I= \displaystyle {\int}_{\lbrace h \in \mathbb{C} :\vert h \vert \geq c_0 \rbrace}\hspace{-1 cm}  e^{-r \vert h - h_c \vert ^2} dh = \frac{1}{r}Q_1(\sqrt{2 r} \vert h_c \vert, \sqrt{2r} c_0),
\end{align}
where $Q_1$ is the  first order Marcum Q function. Also $I$ can be upper bounded as,
\begin{align}
\nonumber
I \leq \frac{1}{r} \frac{c_0}{c_0-\vert h_c \vert} e^{-r(c_0-\vert h_c \vert)^2}.
\end{align}
\begin{proof}
Let $\theta$ denote the phase angle of $h$ and $\gamma$ denote the absolute value of $h.$

Then the term inside the integral $I$ can be written as, $$e^{-r(\gamma ^2 +\vert h_c \vert ^2)} e^{2 \gamma r \vert h_c \vert \cos(\theta - \phi)},$$
where $\phi =\tan^{-1} \left(\frac{Im(h_c)}{Re(h_c)}\right).$ Transforming $h$ into polar coordinates $(\gamma, \theta),$ the integral $I$ can be written as,
\begin{align}
\label{eqn_temp}
I= {\int}_{\gamma = c_0}^{\infty} \gamma e^{-r(\gamma ^2 +\vert h_c \vert ^2)} \frac{1}{\pi} {\int}_{\theta = -\pi}^{\pi}  e^{2 \gamma r \vert h_c \vert \cos(\theta - \phi)} d\theta d\gamma. 
\end{align}

Let $I_0$ denote the Bessel function of zeroth order of the first kind, i.e.,

\begin{align*}
I_0(x)=\frac{1}{2\pi} \int_{-\pi}^{\pi} e^{-x \sin(\theta)} d\theta.
\end{align*}  

Using the transformation $x= \sqrt{2r} \gamma,$ the integral $I$ given in \eqref{eqn_temp} can be written as,
\begin{align*}
I &=\frac{1}{r} {\int}_{\sqrt{2r}c_0}^{\infty} x e^{-\frac{1}{2}\left(x^2+\left(\sqrt{2r}\vert h_c \vert \right)^2 \right)} I_0[\sqrt{2r}\vert h_c \vert x] dx\\
&=\frac{1}{r}Q_1(\sqrt{2 r} \vert h_c \vert, \sqrt{2r} c_0)\\
& \leq \frac{1}{r} \frac{c_0}{c_0-\vert h_c \vert} e^{-r(c_0-\vert h_c \vert)^2}.
\end{align*}
The above inequality follows from the fact that $Q_1(\alpha,\beta) \leq \frac{\beta}{\beta-\alpha}e^{-\frac{(\beta-\alpha)^2}{2}}$ \cite{SiAl}. 
\end{proof}
\end{lemma}
\section*{PROOF OF THEOREM 1}
 Consider the integral $I(H_A)$ given in \eqref{Int_HA_1} (the equations \eqref{Int_HA_1}-\eqref{proof_start5} are shown at the top of the previous page). Substituting the Rician probability density function $f(H_B)$ in \eqref{Int_HA_1} and upper-bounding $Q(x)$ by $e^{-\frac{x^2}{2}},$ we get \eqref{Int_HA_2}. Using the transformation $H'_B=H_B+H_A \frac{\Delta x_A}{\Delta x_B}$ in \eqref{Int_HA_2}, we get \eqref{Int_HA_3}, where $c=H_A \frac{\Delta x_A}{\Delta x_B} + \sqrt{\frac{K}{K+1}}.$

Using Lemma 1, the integral in \eqref{Int_HA_3} can be upper-bounded as given in \eqref{Int_HA_4}. At high SNR, \eqref{Int_HA_4}, can be approximated as in \eqref{Int_HA_5}.

 The upper bound on the probability $P_{ANC}^{MA}{(x_A,x_B)\rightarrow  (x'_A,x'_B)}$ given in \eqref{pe_MA_adaptive_ub} can be written as,
\begin{align}
\label{proof_start}
P_{ANC}^{MA}{(x_A,x_B)\rightarrow  (x'_A,x'_B)}=\int_{H_A \in \mathbb{C}} I(H_A) f(H_A) d H_A.
\end{align}

Substituting the Rician pdf $f(H_A),$ using \eqref{Int_HA_5}, \eqref{proof_start} can be upper bounded as given in \eqref{proof_start2}. Rearranging the terms in \eqref{proof_start2} one gets \eqref{proof_start3}, where $s=-\frac{\Delta x_A}{\Delta x_B}.$ Using Lemma 1, the integral in \eqref{proof_start3} can be evaluated as given in \eqref{proof_start4}. Since $Q_1(a,0)=1$ \eqref{proof_start4} can be simplified as in \eqref{proof_start5}. This completes the proof.

\end{appendix}

\end{document}